\documentclass[11pt]{article}

\usepackage{arxiv}

\usepackage{graphicx}
\usepackage{booktabs}
\usepackage{amsmath}
\usepackage{enumitem}
\usepackage{listings}
\usepackage{xcolor}

\definecolor{codebg}{HTML}{F5F5F5}
\newcommand{\console}{\textsc{console}}

\title{Design and Operation of a Federated GPU Cluster\\for Digital Humanities within DHinfra.at}

\author{%
  Florian Atzenhofer-Baumgartner\thanks{Corresponding author: \texttt{atzenhofer@acm.org}.}\,$^{a}$\quad
  David Fleischhacker\,$^{a}$\quad
  Max Resch\,$^{b}$\quad\\
  Lukas Waldhofer\,$^{a}$\quad
  Michael Otto\,$^{a}$ \\[0.6em]
  \small $^{a}$Department of Digital Humanities, University of Graz, Austria \\
  \small $^{b}$Center for Cultures and Technologies of Collecting, University for Continuing Education Krems, Austria
}
\date{\today}

\runningtitle{Design and Operation of a Federated GPU Cluster for Digital Humanities}

\begin{document}
\maketitle

\begin{abstract}
We describe the design, implementation, and operation of a small federated GPU
cluster built for Digital Humanities (DH) research within the Austrian DHinfra.at
project. The system spans two university sites, brokers logins from a national
identity federation, and exposes compute through three interfaces: interactive
notebooks, SSH, and an OpenAI-compatible inference API. The core of the platform is
a control plane that maps federated identity, per-project quotas, and a model
catalogue onto self-service interfaces. We document this control plane in detail,
including the request-and-approval queue that mediates every privileged cluster
effect and the two-tier model-serving setup, which combines always-on services with
on-demand model swapping and lets researchers consume large language models over
HTTPS without holding a cluster account. We report the constraints that shaped the
build: a fixed power budget, EU-wide procurement, a team of about 1.5 full-time
equivalents, and the relationship between prioritized access and utilization. We
also record the main build decisions against the alternatives we researched and did
not adopt. The platform is released under the Apache License 2.0. We close with a
playbook for teams building a domain-specific GPU cluster on a limited budget. This
is a living technical report, versioned and updated as the platform evolves.
\end{abstract}


\section{Introduction}

GPU compute has become a routine research dependency well outside its original use
in large-scale numerical simulation. Handwritten text recognition, corpus-scale
language modelling, image and video classification, and the everyday use of large
language models (LLMs) now appear in projects whose authors do not identify as HPC
users. These researchers want a model behind a URL, a notebook with a GPU already
attached, and a quota they can read off a dashboard. HPC centres organised around
batch submission, allocation proposals, and queue scheduling serve these
expectations only in part.

DHinfra.at, Digital Humanities Infrastructure Austria, is a 2023--2026
infrastructure project funded by the Austrian Federal Ministry (BMBWF) through the
EU Recovery and Resilience Facility and led by the Department of Digital Humanities
at the University of Graz. It serves a consortium of Austrian institutions with
close ties to CLARIAH-AT, the Austrian node of CLARIN~\cite{hinrichs2014clarin} and
DARIAH. Its remit covers
data-as-a-service, digitisation hardware, and free/open-source software development.
This report concerns its GPU/HPC component: a federated cluster with primary compute
in Graz and secondary compute and storage in Krems. We presented the plans for this
system at ASHPC24~\cite{atzenhofer2024ashpc}; this report documents what was built.

This is the extended version of our ASHPC26 technical-track contribution. The talk
and abstract compress the material; here we give the architecture, the operational
history, and the reasoning behind specific decisions enough space to be reproduced.
We address two readerships: systems and MLOps practitioners interested in the
mechanism, and infrastructure builders in adjacent communities (FAIR data, Digital
Humanities, GLAM) who need to know what a GPU platform requires of them.

\paragraph{Contributions.} This report provides (1) a reproducible architecture for
a small federated GPU cluster that brokers logins from a national identity federation; (2) a
detailed account of the platform's control plane, comprising a request/approval
queue and an inference gateway; (3) an account of operations, covering procurement,
staffing, and the relationship between prioritized access and utilization; (4) a
record of the main build decisions against researched alternatives
(Section~\ref{sec:decisions}), and an Apache-licensed release of the platform; and
(5) a playbook (Section~\ref{sec:playbook}) for building a domain-specific GPU
cluster for LLMs.

\section{Background and Related Work}
\label{sec:related}

We position the system against three areas of research computing and the specific
gap it addresses.

\paragraph{Interactive and accessible HPC.} A long line of work lowers the barrier
between researchers and batch schedulers, from web portals that wrap job
submission~\cite{hudak2018openondemand} to notebook interfaces that spawn compute on
demand~\cite{kluyver2016jupyter}. The underlying resource manager is typically
Slurm~\cite{yoo2003slurm}, with isolation increasingly provided by rootless
containers~\cite{kurtzer2017singularity}. Our interactive tier follows this line. We
extend it by binding the interactive layer to a federated identity broker and a
per-project quota model.

\paragraph{AI/ML on shared and multi-tenant platforms.} The convergence of HPC and
AI has produced large multi-tenant platforms that serve many institutions from
shared, often Kubernetes-based fabric~\cite{weitzel2025nrp}. We reuse the same
multi-tenancy mechanisms, namely scoped credentials, per-tenant accounting, and
heterogeneous hardware, at a smaller scale and with an operational surface sized to
our team and user base.

\paragraph{LLM serving and MLOps.} Efficient LLM inference is a developed systems
area, with paged-attention serving engines~\cite{kwon2023vllm} and mature tooling
for routing, batching, and accounting. Our serving layer reuses these components.
The work specific to our platform is the policy layer around them: per-project token
budgets, model access-control lists, and an OpenAI-compatible gateway through which a
researcher on another institution's virtual machine can consume a model without a
cluster account.

\paragraph{HPC in the humanities.} Despite abundant data, the humanities have been
slow to adopt HPC, for reasons that are organisational and cultural as much as
technical~\cite{terras2018hpc}: workloads are often data-heavy rather than
FLOP-heavy, expertise is thin, and the batch-allocation model fits exploratory,
iterative research poorly. The platform described here is built for these users.
Section~\ref{sec:operations} discusses the resulting emphasis on accessibility and
its effect on utilization.

\section{DHinfra.at: Context and Requirements}
\label{sec:context}

The project moved through four phases: working-group and consortium formation with
governance principles (2023); requirements engineering, training, and
hardware/software research in preparation for procurement (2024); FOSS investment,
acquisition, and test environments (2025); and integration with operations toward a
ready-to-use service, with a public launch targeted for mid-2026.

Three requirements distinguish this setting from a conventional HPC centre and recur
throughout the design.

\begin{enumerate}[leftmargin=1.4em]
  \item \textbf{Federated identity.} Users belong to many institutions and
  authenticate with credentials they already hold, through the national research
  network (ACOnet). There is no separate account application.
  \item \textbf{Self-service configuration.} Quotas, keys, and project setup are
  visible and adjustable through a console. The operational default is prioritized
  access with monitoring.
  \item \textbf{Models served over HTTPS.} Many users want an inference endpoint and
  no shell access. The platform serves models over HTTPS to clients that never
  authenticate to the cluster.
\end{enumerate}

\section{Hardware and Network}
\label{sec:hardware}

The hardware was procured through EU-wide public tendering, which delivered
current-generation accelerators at reasonable cost, comparable to national
``AI Factory'' deployments~\cite{eurohpc_aifactories} and with upgrade pathways.

\begin{table}[ht]
\centering
\small
\begin{tabular}{@{}llll@{}}
\toprule
\textbf{Site} & \textbf{Node} & \textbf{Accelerators} & \textbf{Storage / fabric} \\
\midrule
Graz (primary) & comp02 & 8$\times$ H200 NVL (141\,GB, MIG) & ZFS over 400\,Gbit/s \\
 & comp01 & 4$\times$ H200 NVL + 4$\times$ RTX PRO 6000 (96\,GB) & NDR InfiniBand \\
 & infer01 & 4$\times$ L40S (48\,GB) & (NFS/RDMA, NCCL) \\
\midrule
Krems (secondary) & --- & 6$\times$ H200 NVL (141\,GB) & Ceph; dedicated VMs \\
\bottomrule
\end{tabular}
\caption{GPU inventory across the two sites (current). A further 7$\times$ RTX PRO
6000 Blackwell are on order for Graz. H200 and RTX PRO 6000 are MIG-capable; L40S is
not. Both sites run Ubuntu 24.04 and are intended to converge on a common stack.}
\label{tab:hardware}
\end{table}

The hardware in Table~\ref{tab:hardware} is heterogeneous by design. The H200 NVL
nodes carry large-context and concurrent inference workloads and can be partitioned
with MIG for many small tenants; the L40S and RTX PRO 6000 (Blackwell) cards handle
cheaper, single-tenant exploratory and serving work, and the incoming RTX PRO 6000
expansion roughly triples that pool. This keeps the expensive cards available for
jobs that need them. It also adds complexity to scheduling and model placement: a
model that fits one card's memory may not fit another, and cost accounting reflects
this (Section~\ref{sec:gateway}).

Several constraints shaped the deployment alongside the hardware specification: a
fixed power budget, the provisioning of 400\,Gbit fabric and an ACOnet high-speed
link between the sites, and emergency-power signalling.

\section{Software Stack}
\label{sec:stack}

The stack is built entirely from open-source components. Most are standard tools;
the code we wrote is the integration layer that connects them. We describe the
layers from the bottom up, then the control plane in detail
(Section~\ref{sec:gateway}).

\subsection{Identity and access}
\label{sec:identity}

Authentik\footnote{\url{https://goauthentik.io}} is the identity authority. It brokers ACOnet logins over
SAML and OpenID Connect (OIDC), owns users, groups, and login flows, and exposes an
LDAP outpost
that SSSD consumes to materialise POSIX accounts on cluster nodes. At the edge,
Traefik terminates TLS and applies Authentik \texttt{forwardAuth} as middleware, so
authentication is checked once, centrally, before traffic reaches any service.

Authentik is the sole owner of identity, and no other component duplicates it. Group
membership, leadership, and SSH public keys all live in Authentik; the rest of the
platform reads from it and keeps only a cache, reconciled two ways
(Section~\ref{sec:gateway}). An institution's users appear and disappear in one
place.

\subsection{Scheduling and compute}

Slurm is the resource authority for both user jobs and model-serving jobs, with
SlurmDBD for accounting (backed by MariaDB, which SlurmDBD supports natively, while
the rest of the platform uses Postgres), GPUs tracked as TRES, and pre- and
post-job GPU health checks. SlurmDBD is the authority for job history, which the
control plane does not duplicate.

User containers use the standard rootless HPC split: Enroot\footnote{\url{https://github.com/NVIDIA/enroot}}, a
daemonless runtime that runs entirely as the calling user, and
Pyxis\footnote{\url{https://github.com/NVIDIA/pyxis}}, a Slurm SPANK plugin that adds \texttt{--container-image} to
\texttt{srun}/\texttt{sbatch} and drives Enroot from inside \texttt{slurmstepd}. On the command line, containers are opt-in: a job that names no image runs on
the host. JupyterHub sessions, by contrast, always run inside a container selected at spawn time. Host execution for CLI jobs is acceptable on a semi-trusted research cluster and consistent with the operations stance in Section~\ref{sec:operations}. Curated CUDA/PyTorch images live on NFS as
squashfs files; users layer their own \texttt{pip} packages on top and save the
result. Proxmox VMs and LXC containers, reached over a WireGuard VPN, cover cases
that need stronger isolation or a long-lived dedicated environment.

\subsection{Interactive access}

Researchers reach the cluster three ways: JupyterHub (with a SlurmSpawner, so a
notebook is a Slurm allocation), SSH over VPN, and the inference API
(Section~\ref{sec:inference}). The first two are the standard interactive-HPC
interfaces; the third serves users who do not work from a terminal.

\subsection{The control plane: console, gateway, and approval queue}
\label{sec:gateway}

The control plane is a FastAPI service we call the \console{}. It maps identity,
quotas, and a model catalogue onto self-service interfaces. It presents two faces
over one data model: a public LLM API (bearer-key auth, proxied to the internal
model router) and a session console for members, group leaders, and admins. A
dedicated Postgres database holds budgets, models, usage, API keys, request queues,
an audit log, and the identity cache; Redis holds the auth cache, budget
hot-counters, and a login rate-limit bucket. Background work is minimal: budget
windows are created lazily on first request, usage is written synchronously, and the
only scheduled jobs are an identity reconcile sweep and a project-expiry sweep, both
in-process. No Celery process or additional container is required.

\paragraph{Source of truth and the cache.} The platform assigns one authority per
domain (Table~\ref{tab:authority}). Identity belongs to Authentik, job history to
SlurmDBD, and workspace bytes to ZFS; the \console{} owns LLM budgets, the model
catalogue, usage, and request intent. Identity is mirrored into a local cache and
kept current two ways: an Authentik webhook updates the cache on user and group
changes, and a periodic reconcile pulls each group with its members, repairs drift,
issues exactly one API key per \texttt{(user, group)} pair, and revokes keys for
dropped memberships. A user who disappears from Authentik is flagged \texttt{missing}
and retained; membership workflows then decline to act on the stale target.

\begin{table}[ht]
\centering
\small
\begin{tabular}{@{}lll@{}}
\toprule
\textbf{Domain} & \textbf{Authority} & \textbf{Control-plane role} \\
\midrule
Users, groups, membership & Authentik & cache + API keys + console \\
Job history & SlurmDBD & not duplicated \\
Workspace bytes & ZFS / Ceph & not duplicated \\
LLM budgets, models, usage & \console{} Postgres & full ownership \\
\bottomrule
\end{tabular}
\caption{One authority per domain. The control plane stores approval intent and
derived caches and does not become a second source of truth for identity or storage.}
\label{tab:authority}
\end{table}

\paragraph{The approval queue.} Privileged cluster changes, such as adding a member,
creating a project, granting SSH access, changing a storage quota, or archiving a
project, are not applied by the web process that receives them. The \console{}
records the request as a queued row; a separate worker, \texttt{pm-agent}, polls an
internal endpoint and applies the approved effect on the cluster side (membership,
\texttt{slurm.*}, \texttt{home.*}, \texttt{ssh.grant/revoke}, project
archive/restore/destroy). The split has three consequences. The public-facing service
holds no cluster credentials and performs no privileged side effects; every privileged
action is recorded as a row with an approver and a reason; and the cluster-side worker
can be tested, rate-limited, and reasoned about independently of the web tier.

\paragraph{Request authentication path.} For inference, the path is short and
cache-friendly. A client sends \texttt{Authorization: Bearer <key>} to the public API
host. Traefik applies \texttt{forwardAuth} to an internal verify endpoint, which
returns the resolved user, group, and key (cached in Redis on the key's HMAC for
60\,s) or a 401. Traefik strips any client-supplied identity headers, so the gateway
never trusts identity from outside. The handler resolves the model (an ACL check for
private models), checks the group's daily and weekly token budgets, takes a per-model
concurrency semaphore, and streams to the internal router. Usage is recorded
synchronously on completion. A budget exceeded mid-stream causes the next request to
be rejected; requests already streaming run to completion.

\paragraph{Budgets.} Every budget belongs to a group (a project), never an
individual, in two windows: daily (local midnight) and weekly (ISO week). The cap is
derived at window rollover and frozen for the window, so a user who joins late in a
window receives no extra allowance retroactively:
\[
  \text{cap} = \text{member\_count} \times \text{default\_per\_user\_per\_window}.
\]
Per-group overrides (absolute or a multiplier, set per counter) are a routine
workflow, logged with the author and reason. Each model also carries input and output
cost factors, so a cheap small model can be priced below 1.0 to stretch the same
budget; accounted tokens are raw tokens times the factor. Cost factors let one budget
cover models with very different serving costs.

\subsection{Inference services}
\label{sec:inference}

Model serving is split into two tiers, and the gateway routes and meters requests
without running models itself.

\begin{itemize}[leftmargin=1.4em]
  \item \textbf{Always-on tier:} a set of open-weight models stays resident, each holding its GPUs through a long-lived Slurm allocation, behind the internal router, exposing OpenAI-compatible chat, completion, and embedding endpoints with streaming, including multimodal (image) input. Footprints vary with model size and quantization: a quantized small model serves from a single L40S, while the largest
  resident models occupy a full eight-H200 node each. Partner-institution
  researchers call these endpoints with a scoped key and need no cluster account;
  this is the most-used interface in practice.
  \item \textbf{On-demand tier:} models outside the resident set load on first
  request and unload to free their GPUs when idle --- the long tail of models too
  large or too rarely used to keep resident, running across multiple H200s and
  optionally pinned to a node. We validated on-demand swapping at a DHd2026 workshop
  with concurrent heavy users and teacher/student model swapping on shared H200
  nodes.
\end{itemize}

The router (llama-swap\footnote{\url{https://github.com/mostlygeek/llama-swap}}) holds no GPUs; it is an HTTP front for engine lifecycle.
Each model's launch command allocates GPUs as a Slurm job under a single trusted
service account and runs the engine --- vLLM~\cite{kwon2023vllm}, or llama.cpp\footnote{\url{https://github.com/ggml-org/llama.cpp}} for
one-slot models --- in a container on the assigned node (the \texttt{salloc}
$\rightarrow$ \texttt{srun} $\rightarrow$ \texttt{docker run} chain, against the root
Docker daemon, separate from the unprivileged Enroot path used for user jobs). A GPU
is therefore held only while its model is loaded. Always-on versus on-demand is a
per-model loading policy: resident models are exempt from idle unloading, and Slurm,
not the router, arbitrates the GPUs between them. Tensor-parallel width, partition
(L40S vs H200), and optional node pinning are set per model; cold loads allow a
long health-check window because a first load means a weight download plus
CUDA-graph compilation.

The gateway's controls over this layer are deliberately small: a per-model
concurrency limit (a bounded semaphore --- generous for batched vLLM, one or two for
single-slot llama.cpp), the public/private visibility and per-user ACL of each
catalogue entry, and the input/output cost factors used for budget accounting
(Section~\ref{sec:gateway}). The roadmap integrates both tiers under one controller
with project management and asynchronous batch scheduling.

\subsection{Monitoring and configuration}

Prometheus and Grafana with DCGM exporters provide GPU and node observability. The
metrics surface inefficient resource use, which we raise with the relevant users.
Platform services run as Docker Compose plus systemd on the head node. We are also
evaluating NixOS for declarative node configuration, with a single description,
atomic updates within a release, and per-user environments; it is currently under
test for the monitoring and identity components, and both sites are intended to
converge on the same stack.

\subsection{Local testbed and reproducibility}
\label{sec:testbed}

A local testbed reproduces the identity, provisioning, and gateway chain in Docker
Compose, so the platform can be exercised from a cold start without cluster
hardware. Dex\footnote{\url{https://github.com/dexidp/dex}} stands in for the institutional SSO, a mock model router replaces the
GPU serving path, and the provisioning worker runs against mock ZFS and Slurm
backends. The mock/live boundary is one switch per backend, and the component
contract is identical on both sides. A test harness brings the stack up, seeds
projects, and runs end-to-end checks over the real HTTP surfaces: identity
reconcile, the approval queue, key issuance, and budget enforcement. The result is a
control plane that is testable in continuous integration, and a contributor can
validate a change before it reaches the cluster.

\subsection{Design decisions and rejected alternatives}
\label{sec:decisions}

Several components were chosen against a credible alternative that we researched and
did not adopt. Table~\ref{tab:decisions} records the main ones with the deciding
reason.

\begin{table}[ht]
\centering
\small
\begin{tabular}{@{}p{0.20\linewidth}p{0.22\linewidth}p{0.20\linewidth}p{0.30\linewidth}@{}}
\toprule
\textbf{Concern} & \textbf{Chosen} & \textbf{Not adopted} & \textbf{Deciding reason} \\
\midrule
Service orchestration & Docker Compose + systemd & Kubernetes / K3s &
About ten services on a handful of nodes; Compose is sufficient and simpler. K3s is
a growth path past roughly fifteen services or per-user pod isolation. \\
User containers & Enroot + Pyxis (rootless, SPANK) & Docker for user jobs &
Docker's root daemon is equivalent to root on the node; unsafe for multi-tenant user
containers. \\
Model-weight storage & Served from NFS over RDMA & Local NVMe cache on compute nodes &
Benchmarked: the head node's 22-drive ZFS pool over InfiniBand beats a compute node's
two-drive local NVMe (140\,GB cold start ${\sim}5$\,s vs ${\sim}20$\,s). \\
Identity host & Head node or dedicated VM & External VPS over a tunnel &
A tunnel outage longer than the SSSD cache TTL stales all identity and blocks new
logins. \\
Slurm accounting DB & MariaDB & Postgres (used elsewhere) &
SlurmDBD supports MySQL/MariaDB natively; Postgres is not a first-class backend. \\
LDAP outpost mode & Direct search & Cached &
Cached mode delays membership propagation by minutes to hours. \\
Inference admission & Gateway does ACL, budget, and metering & Generic external router &
Keeps admission policy and accounting in one service rather than split across a
router and a sidecar. \\
\bottomrule
\end{tabular}
\caption{Main build decisions and the alternatives considered. Reasons are specific
to this deployment's scale and team; a larger site might decide differently.}
\label{tab:decisions}
\end{table}

\section{Operations}
\label{sec:operations}

The operating model reflects a small team, roughly four to five core developers at
about 1.5 full-time equivalents, and a community-oriented approach. The clusters run
in dedicated networks, effectively a DMZ, independent of central university HPC while
using institutional network, identity federation, and facilities. The stated
priorities are simplicity, security, and maintainability. Infrastructure-as-code is a
goal that is partially realised.

\paragraph{Prioritized access and utilization.} The platform grants prioritized
access and accepts lower peak utilization as a result. Early operation showed the
expected effect: prioritized access exposes optimisation gaps in user code, and some
resource use is inefficient. We respond by monitoring usage and contacting the users
involved.

\paragraph{Procurement and tendering.} EU-wide public procurement delivered
state-of-the-art hardware at reasonable cost, at the price of substantial process
overhead. Offer quality varied widely, one site's tender required a second round, and
the practical lessons were straightforward: vet bidders, keep direct contact, and
involve the legal department early. Specifications and governance were re-negotiated
after the proposal was accepted, and teams should plan for this.

\paragraph{Concurrent development.} Much of the effort was concurrent: hardware and
software developed together, technical and policy work developed together, and the
partner institutions moved at different paces. For an infrastructure project with
limited personnel funding, a large share of the cost is staff training. Investment in
staff and in detailed specifications repaid itself: it built in-house knowledge,
reduced downtime, and lowered dependence on external experts for intricate problems.

\section{Use Cases}
\label{sec:usecases}

A representative DH pipeline exercises every tier in turn. (1) In an exploratory
phase the researcher tests models interactively on JupyterHub, varying architectures,
sizes, quantizations, and prompts, and benchmarks them on a domain task such as
information extraction. (2) They select a model on the evidence. (3) The model is
deployed through the controller as an API or a batch job; a project running on
another institution's VM then calls the DHinfra inference endpoint over a whitelisted
HTTPS route. (4) Secondary services, such as project storage or custom VMs at Krems,
support the work. (5) Results feed downstream into research outputs.

The pilots cover a range of workloads: automatic text recognition, image and video
classification, VR applications, and adaptive AI tutoring, alongside DH projects on
historical job ads, digitised newspapers, archival registers, and diplomatics. The
scheduling and serving policies must accommodate this range.

\section{A Playbook for a Domain-Specific GPU Cluster}
\label{sec:playbook}

This section distils the build into guidance for a team standing up a GPU cluster for
LLMs and DH/GLAM-style workloads on a modest budget. We organise it around complexity
tiers: start with the simplest robust option and add components when a workload
requires them.

\paragraph{1. Decide whether you deliver nodes or inference endpoints.} If your users
are HPC-native, give them Slurm directly. If they are not, as in the DH and GLAM case,
the primary deliverable is an inference endpoint with a quota, and the cluster sits
behind it. This choice determines the rest of the design.

\paragraph{2. Federate identity and keep a single authority.} Run one identity
authority (we used Authentik over SAML/OIDC against the national federation) and let
other components hold only a reconciled cache. Assign one authority per domain
(Table~\ref{tab:authority}); this is what keeps a federated system debuggable.

\paragraph{3. Separate intent from effect.} Do not let a web request change cluster
state directly. Record the request, approve it, and have a separate worker that holds
the credentials apply it. This boundary provides auditability, least privilege, and a
cluster-side surface that can be tested on its own.

\paragraph{4. Meter consumption per project.} Attach quotas to projects, account in
tokens with per-model cost factors so that cheap models stretch a budget, and freeze
caps for the duration of each window. Treat overrides as a routine, logged workflow.

\paragraph{5. Serve models in two tiers.} Run an always-on tier (a batched serving
engine such as vLLM behind a router) for steady demand and an on-demand swapping tier
for the long tail. Keep the gateway separate from the GPU path: it routes and meters
requests, and the scheduler runs the serving engine.

\paragraph{6. Account for heterogeneous hardware.} Mixed GPUs, with large cards for
concurrency and smaller cards for exploration, lower cost, provided the accounting and
placement track which card serves a given model.

\paragraph{7. Set an explicit utilization target.} A domain platform can accept lower
utilization in exchange for accessibility and fast iteration. State the chosen target
explicitly and support it with monitoring and direct user contact.

\paragraph{8. Make the control plane testable without the cluster.} Mock the
privileged backends (identity, storage, scheduler) behind the same interface the
live ones use, and run the full request and approval chain in containers. A
cold-start testbed lets you validate control-plane changes in continuous integration
before they reach the hardware.

\paragraph{9. Budget for staff and specifications.} EU-wide procurement can deliver
excellent hardware but requires time, legal support, and detailed specifications. The
staff who write a good tender are the staff who operate the cluster afterwards. Expect
to re-negotiate specifications and governance after funding is approved.

\section{Open-source release}
\label{sec:release}

The platform is released under the Apache License 2.0 and will be public at the
launch; the cluster-side configuration and operator runbooks are published alongside
it. Capable multi-tenant GPU and LLM-serving stacks already exist, but equivalent
functionality is commonly available only under enterprise licensing or as a managed
cloud product. A permissively licensed, self-hostable stack aimed at small and
mid-size domain-specific deployments fills a gap in that space: another institution
can run the same identity-federated, quota-metered inference platform on its own
hardware, with no per-seat or per-GPU licensing. The components the platform
integrates (Slurm, Authentik, Enroot/Pyxis, vLLM, Traefik) are themselves
open-source, so the contributed part is the control plane and the integration that
binds them.

\section{Open Questions and Outlook}
\label{sec:outlook}

Three questions remain open, and we bring them to the systems and research-computing
communities. The first concerns the relationship between federated identity and
institutional quotas. We keep a single identity authority and derive per-project
budgets from it, but a national federation spans institutions that may each want to
set and account for quotas over their own members. How should a platform reconcile
fine-grained, per-institution accounting with one authoritative identity source,
without letting each institution become a second place where membership and
entitlements are defined? We do not have a clean answer, and the tension grows as the
user base widens across the consortium.

The second concerns caching on heterogeneous GPUs. The same model may be served from
cards with quite different memory profiles, so the working set a KV cache or a
weight cache can hold changes from one placement to the next. What is the right
caching and cache-admission strategy when a model's effective capacity depends on the
card it is placed on, and how much of that placement should the scheduler be told about
rather than leaving it to each serving engine? We suspect the answer is largely
platform-specific, but a more general treatment would help.

The third concerns mixed scheduling. Latency-sensitive inference and
throughput-oriented batch work compete for the same pool, and the two have opposite
tolerances for queueing delay. How can a scheduler interleave them on shared hardware
so that interactive requests stay responsive without leaving accelerators idle
between batch jobs? Fair-share accounting answers part of this, but the interaction
between admission control at the gateway and job scheduling in Slurm is not something
we have resolved.

Our current work expands the user base to the full CLARIAH-AT consortium, deepens Krems
integration (distributed storage and coordinated, tracked GPU use), maintains a CLI-
and API-first interface with a GUI as a secondary layer, and targets a public launch
the second half of 2026. Two trends are converging: general HPC is broadening access, and
domain-specific systems are streamlining particular use cases. Small-scale,
domain-specific HPC complements large-scale general HPC.

\section{Conclusion}

Most of the engineering effort in this platform went into the control plane. The
accelerators and network fabric are standard hardware. The cluster becomes usable by
researchers who are not HPC users through the layer that federates their identity,
meters consumption per project, and serves models over HTTPS, with a separation
between recording a request and applying its privileged effect. We have described this
layer in enough detail to reproduce, and we have stated the operational trade-offs in
accessibility, utilization, and staffing that a small team has to decide explicitly.
As a living document, this report will be updated as the platform moves to public
operation.

\section*{Acknowledgements}

DHinfra is funded 2023--2026 by the Austrian Federal Ministry of Education, Science and
Research (BMBWF) under ``(Digitale) Forschungsinfrastruktur'', in cooperation with the EU
Recovery and Resilience Facility (RRF), NextGenerationEU. We thank EuroCC, VSC, and ASC for guidance;
the ASHPC community for feedback at and since ASHPC; and the DHinfra core team and the
IT Services of the University of Graz and the University for Continuing Education
Krems. We acknowledge an ASHPC26 scholarship.

\bibliographystyle{plain}
\bibliography{references}

\end{document}